\documentclass[journal]{rmaa}
\usepackage{psfrag,color}

\usepackage[latin1]{inputenc}

\newcommand{\ggs}{Garc\'{\i}a-Segura }

\title{The 3D velocity structure of the planetary nebula NGC~7009}

\author{
  W. Steffen,\altaffilmark{1}
  M. Esp\'{\i}ndola,\altaffilmark{2}
  S. Mart\'{\i}nez\altaffilmark{3}
   and
  N. Koning,\altaffilmark{4}}

\altaffiltext{1}{Instituto de Astronom\'\i{}a  UNAM, M\'exico.}
\altaffiltext{2}{Instituto Polit\'ecnico Nacional, M\'exico.}
\altaffiltext{3}{Benem\'erita Universidad Aut\'onoma de Puebla, M\'exico.}
\altaffiltext{4}{Space Astronomy Laboratory, University of Calgary, Canada.}

\shortauthor{Steffen, Esp\'{\i}ndola, Mart\'{\i}nez \& Koning}
\shorttitle{The 3D velocity structure of the planetary nebula NGC~7009}

\fulladdresses{
\item Wolfgang Steffen: Instituto de Astronom\'{\i}a,
Universidad Nacional Aut\'onoma de M\'exico,
Apartado Postal 106, 22800 Ensenada, B.C., M\'exico
  (wsteffen@astrosen.unam.mx).
}

\listofauthors{W. Steffen,  Esp\'{\i}ndola M., Mart\'{\i}nez, S. \& N. Koning}
\indexauthor{Steffen, W.}
\indexauthor{Esp\'{\i}ndola M.}
\indexauthor{Mart\'{\i}nez, S.}
\indexauthor{Koning, N.}

\abstract{In search for deviations from homologous expansion in planetary nebulae
we present a 3D morphokinematical model of NGC~7009. The model has been constructed with {\em Shape} based on PV diagrams from the literature and HST
images. We find that the data are consistent with a radial velocity field with increased gradient at high latitudes compared to the equatorial region (Model 1). In a second model we assume a linearly increasing radial velocity component with an added poloidal component of order 10~km/s at latitudes around $70^\circ$. The true velocity field is likely to be in between these two limiting cases. We also find that the expansion of the ansae is non-radial with reference to the central star. Their velocity field is focused near the apparent exit points from the main shell. We predict the proper motion pattern for the model with a non-zero poloidal velocity component.    }

\resumen{En busca de desviaciones de una expansi\'on hom\'ologa en nebulosas planetarias
presentamos un modelo 3D morfo-cinem\'atico de NGC~7009. El modelo ha
sido construido con {\em Shape} basado en diagramas posici\'on-velocidad
de la literatura e im\'agenes del HST. Encontramos que los datos son
consistentes con un perfil de velocidad radial con un gradiente mayor a latitudes comparado
con el de la regi\'on ecuatorial (Modelo 1). En un segundo modelo supusimos una componente
de velocidad radial que incrementa de manera lineal con una componente poloidal adicional del
orden de 10~km/s a latitudes alrededor de $70^\circ$. El verdadero campo de velocidad probablemente
es intermedio entre estos casos l\'{\i}mites. Encontramos que la expansi\'on
de los {\em ansae} no es radial con respecto a la estrella central. Su campo de velocidad parece
expandir de cerca del punto de salida de la c\'ascara principal. Predecimos un padr\'on de
movimiento propio para el Modelo 2.
}

\addkeyword{Planetary Nebulae: Individual: NGC7009}
\addkeyword{ISM: Outflows}
\addkeyword{Stars: Post-AGB}
\addkeyword{Stars: Mass loss}

\begin{document}
% Typeset article header
\maketitle

\section{Introduction}
\label{sec:intro}

In recent years it has become clear that a complex three-dimensional (3D) structure of a planetary nebula (PN) poses a major challenge to the interpretation of its formation process. As has been shown for NGC~7009 by Gon{\c c}alves et al. (2006), even the interpretation of spectroscopic measurements for abundance determinations may be seriously affected by the 3D structure of an object. Furthermore expansion parallax determinations of badly needed improved distances to PNe are dependent on the accuracy of 3D models of their structure and velocity field (e.g. Terzian 1997; O'Dell et al., 2009). It is well known that orientation effects are a problem for any structural classification scheme or statistical studies of planetary nebulae (e.g. Balick \& Frank, 2002). It is therefore crucial to obtain 3D morphokinematical information and reconstructions.

NGC~7009 is a planetary nebula with several morphological and kinematical sub-systems with multiple shells, a halo, jet-like streams, ansae and small-scale filaments and knots. Sabbadin et al. (2004; STC04, for short) have produced a tomographic 3D-reconstruction of NGC~7009 based on long-slit observations at the ESO NTT telescope with the EMMI instrument. Their reconstruction is based on the assumption of a linear and radial, i.e. homologous, velocity field.

Homologous expansion is evidenced in axisymmetric objects that show the same structure in direct images and in their echellograms (or position-velocity, P-V, diagram). The P-V diagrams encode the velocity along the line of sight due to Doppler-shift of the spectral line. In homologous expansion the Doppler-shift corresponds directly to the position along the line of sight, except for a constant factor. In axisymmetric objects this factor can be determined by fitting an axisymmetric morphokinematic model simultaneously to the image and PV diagrams. If such a constant factor
can not be found to fit the whole image and PV diagrams, then deviations from a homologous expansion are present.
These might be in the form of a non-linear dependence of the radial velocity (radially outward)
as a function of distance from the central star or a non-zero poloidal velocity component.
Deviations from a linear correspondence between image and PV diagram may, however, also be due to structural deviations from
axisymmetry. In such a case, if the velocity field is known and monotonous in its components, the 3D emission structure can be inferred unambiguously from the image and P-V diagrams. However, if the velocity field is not well
known and the object is intrinsically 3D, then ambiguities arise in the determination of the true 3D structure.

Hydrodynamical simulations show that in wind-driven ellipsoidal and bipolar nebulae, deviations from a homologous expansion may introduce substantial distortions in reconstructions of the 3D structure that assume homologous expansion (Steffen et al., 2009). We are therefore undertaking a study of well-resolved planetary nebulae with detailed imaging and internal velocity measurements (using Doppler-effect and/or internal proper motion) in order obtain 3D reconstructions that take into account such deviations. In this paper we analyze the well-observed nebula NGC~7009 in search of morphokinematical solutions that might provide information about the dynamical state of the nebula.

STC04 find that the large scale structure of the main shell is best described as a triaxial ellipsoid. Guerrero,  Gruendl \& Chu (2002) find extended X-ray emission which precisely fills the main shell. Since the shocked fast stellar wind is expected
to produce such x-ray emission, this observation supports the existence of a wind
or at least a high pressure bubble that drives the expansion.
Steffen et al. (2009) showed that morphokinematical reconstructions of such nebulae, when based on a homologous expansion law lead to an overestimation of the bubble cross section along the line of sight at mid to high latitudes. In the reconstructions by STC04, such a difference in the cross sections along the line of sight and in the plane of the sky is clearly present.

In this paper we therefore estimate the deviations from homologous expansion that are still consistent with the current imaging and kinematical data while maintaining approximately the same cross section of the main shell in the plane of the sky and along the line of sight. We construct a model with purely radial velocities but increasing velocity gradient with distance (Model 1) and a linearly increasing radial component with an additional poloidal component. With these boundary conditions the models should provide upper limits for these velocity components. In order to allow testing of these results we predict internal proper motion patterns to be compared with future observations.

Fern\'andez, Monteiro and Schwarz (2004) provided proper motion measurements of the eastern ansa in NGC~7009 that revised the values obtained by Liller (1965). They also detected expansion of the main shell, but did not quantify it, since they suggested that it might be a moving ionization front, which does not provide the velocity of the bulk motion of the gas. However, STC04 find that the nebula is fully ionized, except possibly in the ansae and the caps, such that the concern about a misinterpretation of the measured velocities might not be justified. Rodr\'{\i}guez \& G\'omez (2007) have recently used radio observations with the Very Large Array (VLA) to determine the tangential velocity of ansae and find that the western ansa might be faster than the eastern one by a facter of 1.5. However, the error bounds of the two values overlap, such that a similar velocity for both is not excluded (especially considering the fact that their distance to the central star is different by less than ten percent). Based on our models we predict proper motion patterns of the main shell and the ansae, which are crucial to eliminate or reduce ambiguities in future models.

\begin{figure}[!t]\centering
  \vspace{0pt}
  \includegraphics[angle=00,width=0.85\columnwidth]{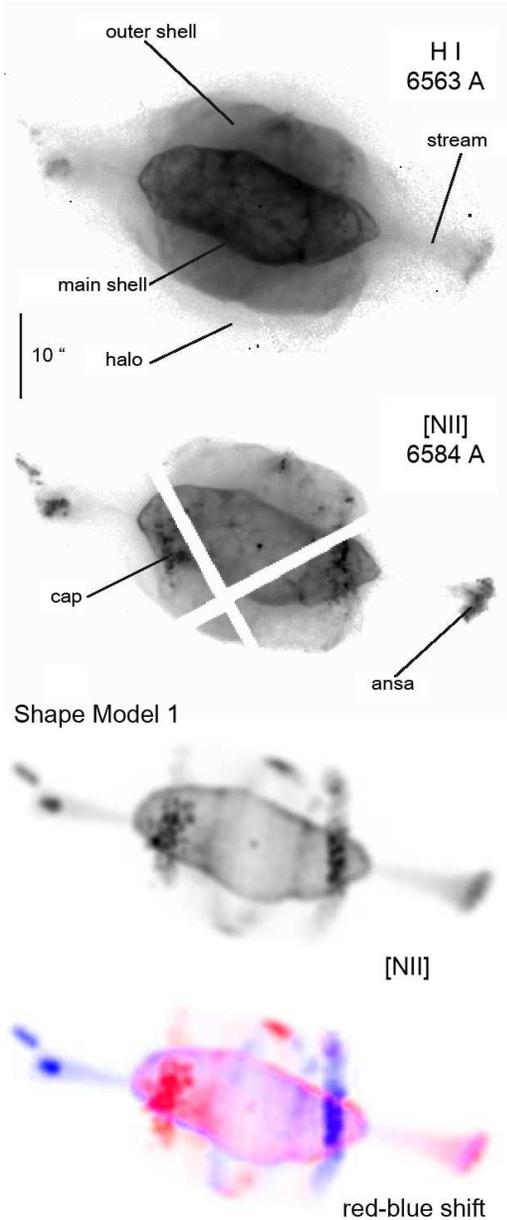}
  \caption{The first and second image show NGC~7009 are HST images in H$\alpha$
    and [NII] adapted from Sabbadin et al. (2004). The third and fourth image are
    our {\it Shape} model based on [NII] position-velocity diagrams from
    the same paper and HST images (see Figures 3 and 4).
    The bottom image has been rendered with red and
    blue color coding according to the Doppler-shift.
    The spatial resolution in the model images is 0.5 arcsec}
  \label{fig1}
  \vspace{0pt}
  \end{figure}

 The tomographic reconstruction by STC04 uses ground-based position-velocity diagrams at regular intervals of 12 position angles. Although these are of excellent quality, the coverage of the object is not complete and there are gaps between slits which had to be interpolated. Furthermore, the spatial resolution is seeing limited. In this work we therefore intend to improve on the spatial resolution and fill in gaps by taking into account high-resolution imaging information from the Hubble Space Telescope in a new, manual, reconstruction with our morphokinematic modeling package {\it Shape}.

\begin{figure}[!t]\centering
  \vspace{0pt}
  \includegraphics[angle=180,width=0.95\columnwidth]{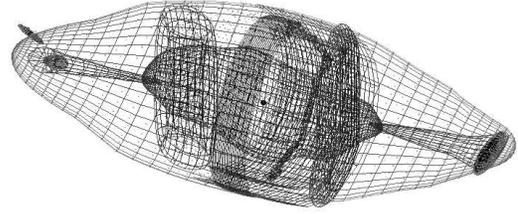}
  \caption{The mesh representation of our 3D-model of NGC~7009.
  These meshes
enclose the volumes that correspond to the various features of the {\it
Shape}-model. The volumes are sampled with randomly distributed particles. Highly
tunable 3D-functions
are then used to quantify the relative emission and velocity fields.
}
  \label{fig2}
  \vspace{0pt}
  \end{figure}

The layout of our paper is as follows. In section 2 we describe the modeling procedures and section 3 contains the results and discussions for the main structures of NGC~7009. In section 4 we summarize our conclusions.

\section{Morphokinematic modeling with SHAPE}
\label{sec:shape}

We have modeled the morphokinematical structure of NGC~7009 with the {\it Shape}
software (version 2.7)  which is
a rather new approach to reconstructing the 3D structure of astrophysical objects from image and Doppler-data (Steffen \& L\'opez, 2006; Steffen, et al., 2009).
The model consists of a mesh-structure for each significant large-scale and small-scale structure that has been identified in the image data (Figure \ref{fig1}). The meshes have been constructed
with the modeling tools that {\it Shape} provides for this purpose. Figure (\ref{fig2}) shows the resulting 3D mesh with the orientation that corresponds to Figure (\ref{fig1}).

Note that the reconstruction of the 3D structure based on 2D images and PV diagrams is not a unique process and relies on assumptions that allow us to map the Doppler-velocity to a spatial position along the line of sight. The assumptions can involve the structure of the object, the velocity field or both. Often some kind of symmetry is assumed to reduce ambiguities. Magnor et al. (2005) have developed an algorithm to reconstruct nebulae with a cylindrical symmetry that relies on 2D images only. The assumption of a homologous expansion, i.e. a velocity vector that is proportional to the position vector, allows limited reconstruction of arbitrary spatial distributions based on Doppler velocity measurements to within a constant of proportionality along the line of sight. If the extent of the object along the line of sight can be inferred from other sources, e.g. large scale symmetry, then this reconstruction is unique. The reconstruction of NGC~7009 by STC04 is of this type.

Many planetary nebulae appear to show a homologous expansion as evidenced by the similarity of structures
in their images and PV diagrams.
However, we would like to point out that caution has to be exercised when relating the homologous
expansion of a highly non-spherical shell-like structure with velocity fields in hydrodynamical simulations. The high density shells represent the same hydrodynamical or ionization feature in different directions as seen from the central star. Such a feature could be an ionization front or the cooling shock region caused by the fast stellar wind impacting on the slow AGB-wind. In a spherically symmetric object we then see a spherical shell. This shell is of course not empty inside and is surrounded by external material. Along each direction from the central star we then have a velocity profile as a function of distance. As shown in detail by hydrodynamical simulations, the velocity profile along such a ``ray'' may be very complex and usually is far from linear (e.g. Sch\"onberner et al., 2007), i.e. along a single ray, the expansion is non-homologous.

However, the story is very different when we consider the {\em same} hydrodynamical feature in different directions. Usually, due to varying ambient densities in different directions, the expansion speed of the same hydrodynamical feature changes with direction. This is the main reason for the non-spherical shapes of planetary nebulae in the generalized interacting stellar wind model (Kahn \& West, 1985; Balick 1987). In a first approximation, the expansion along each direction is independent from the other directions and features are ``sorted'' in a similar fashion as in a ballistic expansion: the faster, the further away, such that a near-homologous velocity field arises {\em for each hydrodynamical or ionization feature}. This means that flow problem admits a similarity solution (e.g. Kahn and West, 1985). However, the pressure is not exactly constant along surfaces of these stratified shells. Therefore non-radial velocity components and deviations from the homologous radial expansion may arise. Two- or three-dimensional numerical hydrodynamical simulations show such deviations from radial motion (e.g. Mellema et al., 1991; Steffen et al., 2009). The simulations exhibit deviations from a linear variation of the radial velocity component with distance as well as non-zero poloidal velocity components. Fortunately, except for local small-scale deviations, the radial component increases monotonically with distance and the poloidal
component varies rather smoothly with poloidal angle. This behavior reduces considerably the ambiguities inherent in
the general problem of mapping the Doppler-velocity to a position along the line of sight.

The classification and characterization of the velocity field of observed planetary nebulae is usually done based on such hydrodynamical or ionization shells. In order to judge with any degree of reliability, whether the velocity field is homologous or not, only such coherent structures can be used to compare their structure in images and PV diagrams of a single emission line.

In contrast, to obtain a velocity field that can be compared with one-dimensional simulations, the velocity field {\em across} such shells has to be determined. To some extend this can be achieved considering different ionic species which trace different regions along any given direction. This type of analysis has been done, for example, by Wilson (1950) and recently by Sabbadin et al. (2004 and references therein).

An important fact to take into account here is that the velocity as a function of distance (measured in different directions) of a single non-spherical shell-like feature is expected to differ from that measured along a single radial direction (which crosses all hydrodynamical structures). Hence, the 3D velocity field for different shells may also be very different from each other. If different ionic species occupy distinct spatial regions, their velocity fields with distance and poloidal angle can not be assumed to be the same. For a reliable 3D reconstruction of an object it is therefore very important to distinguish these different shells and reconstruct or model them independently.
{\em Shape} is excellently suited for this kind of analysis, since it allows the user to judge which features to
reconstruct independently, even in a single line observation.

Our main aim is to obtain information on the deviations of the velocity field from
homologous expansion within these coherent shell structures.

Instead of making a strong assumption about the velocity field, we make an assumption about the structure along the line of sight. The main assumption is that the cross section along the line of sight should be the same or very similar to the cross section in the plane of the sky at the same position along the axis of the main shell.
A secondary, but useful constraint is that the streams should touch the tips of the main shell, both in the plane of the sky as well as along the line of sight.

In our model, velocity components are separable, i.e. the poloidal velocity is only
dependent on the "latitude" and the radial velocity is only dependent on distance.
This introduces a further constraint by reducing the degrees of freedom
but relaxes the constraint of a homologous expansion.

The kinematics of the outer shell indicates a systematic blue-shift by approximately 2-3 km s$^{-1}$ with respect
to the main shell. The origin of such a shift can be either structural, i.e. it is closer to the observer. It
could be truly slower, due interaction with a density gradient or proper motion in the ambient medium. The effect
is small and has been neglected in our model, which is concerned mostly with the kinematics of the main shell.
Any effects on the main shell would be unnoticeable.

We present two solutions that are qualitatively different in terms of their velocity field. In Model 1 we have a radial velocity field with a monotonous non-linearly
increasing velocity magnitude. In Model 2 a linearly increasing radial velocity
component and a non-zero poloidal component is applied. The azimuthal component
is always assumed to be zero. These assumptions are based on the results from our
recent study of the velocity fields in hydrodynamical simulations of typical
ellipsoidal and bipolar planetary nebulae (Steffen et al., 2009),
where a single dense hydrodynamical shell has been extracted and analyzed. The simulations always show both, an increasing gradient for the radial component and a non-zero poloidal component at mid latitudes. Cylindrical and mirror symmetry in the simulations demands that the poloidal velocity is zero at the equator and at the axis.

We start with an ellipsoidal structure along the line of sight that is the same as that of the image outline in the plane of the sky. This structure and the velocity field are then iteratively adjusted to conform to the observed PV diagrams. While this procedure does not provide a unique solution, it is expected to provide estimates for the deviations from a homologous expansion. Based on the results we then make predictions for the expansion in the plane of the sky, which can be verified with future proper motion measurements of nebular features.

Our modeling of NGC~7009 is based on images from the Hubble Space Telescope (HST) and position-velocity (PV) diagrams from STC04. Please refer to this paper for details on the observations. Figure (\ref{fig1}) shows the HST [NII] image and our corresponding model image.  Although, for consistency, the images and PV diagrams of additional spectral lines have been consulted, we concentrate on the modeling of the [NII] emission. [NII] shows the highest contrast between features and is more likely to be concentrated along surfaces or in discrete knotty structures. These properties provide a less ambiguous mapping between imaging and spectral features.  Note that the figures in STC04 are displayed with a logarithmic grayscale, i.e. the contrast between features is greatly reduced. In our work we focus on the relative positions and velocities of the structures in NGC~7009, rather than their relative brightness, such that the key information is in the outlines and positions of the features. Currently, {\it Shape} can not reproduce these high contrast structures and therefore the relative brightness should be taken to be only qualitative, intended for identification of the individual features and for visualization purposes.

\begin{figure}[!t]\centering
  \vspace{0pt}
  \includegraphics[angle=-0,width=.85\columnwidth]{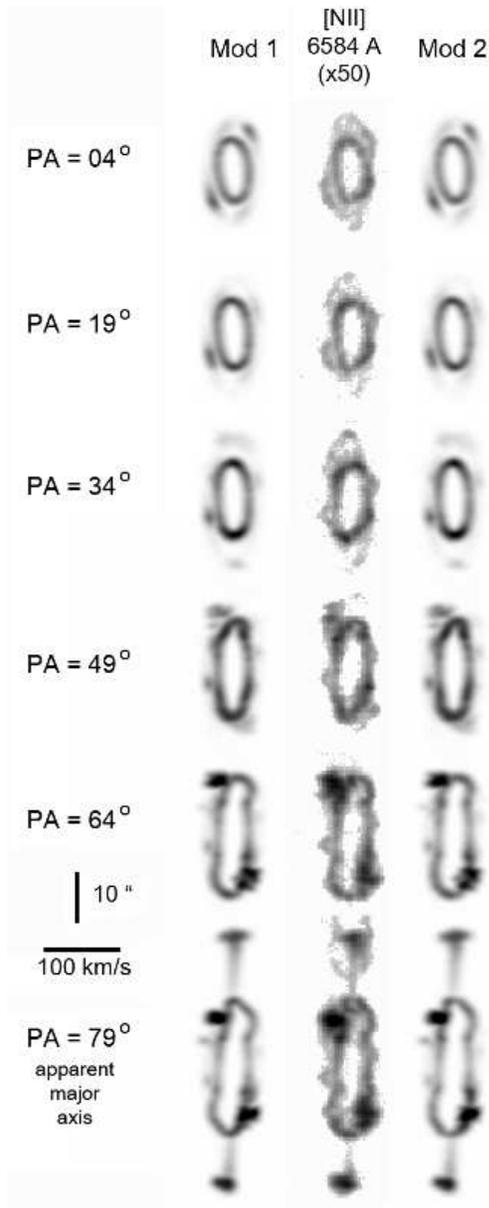}
  \caption{The middle column shows [NII] position-velocity diagrams of
  NGC~7009 at 6 different position angles separated by $15^\circ$
from each other  (adapted from Sabbadin et al., 2004).
The corresponding model PV diagrams are shown to the left (Model 1) and right (Model 2).
Note that the grayscale of the observations is logarithmic and that of the observations
is square-root (see main text). The spatial resolution in the PV diagrams is 0.5 arcsec and the
velocity resolution is 5 km/s.}
  \label{fig3}
  \vspace{0pt}
\end{figure}

\begin{figure}[!t]\centering
  \vspace{0pt}
  \includegraphics[angle=-0,width=1.0\columnwidth]{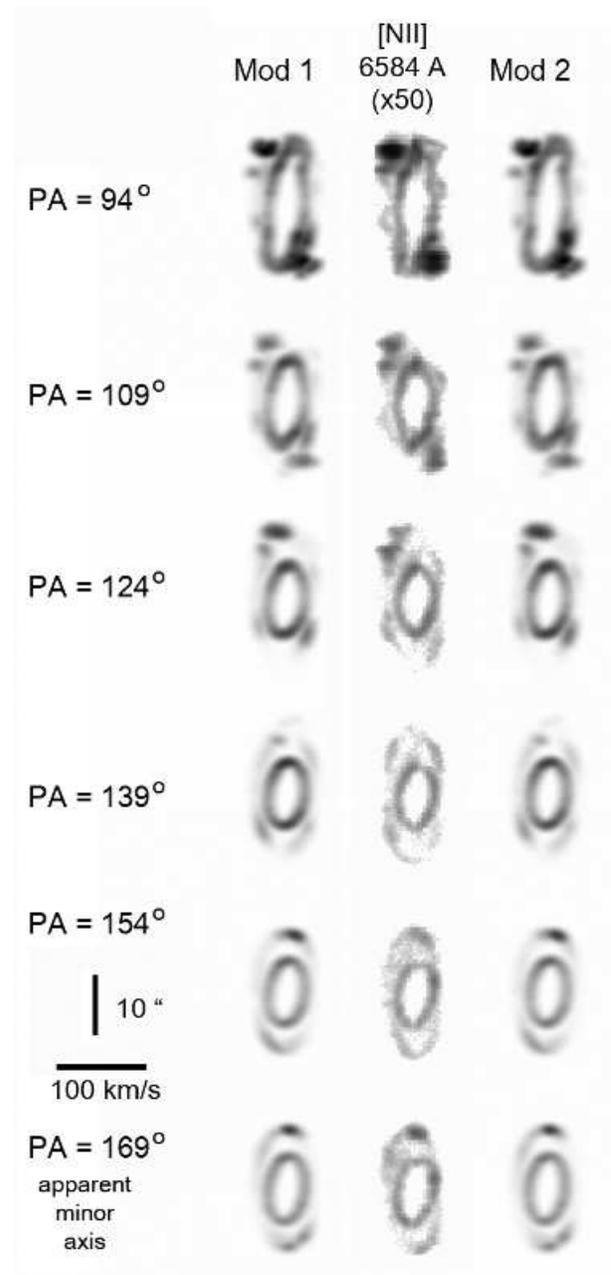}
  \caption{Same as Figure (3), but for 6 additional position angles.}
  \label{fig4}
  \vspace{0pt}
\end{figure}

The general modeling strategy is as follows.
From the overall characteristics of the images (Figure \ref{fig1}) and PV diagrams (Figure \ref{fig3} \& \ref{fig4}, adapted from STC04), as well as previous models and reconstructions, a first approximation of the model is created by interactively constructing the mesh structure. This mesh structure is filled with a random distribution of particles (see below). Initially, a homologous expansion is assumed for the velocity field, according to the results from
STC04. Then the general orientation of these structures is changed
to fit the tilt of the PV diagrams. For the main and the outer shell we first assume that the cross-section perpendicular to the axis is circular. We then apply large scale first order corrections to the shape and velocity field. These corrections include triaxial structure, bending and rotation around the axis. For the velocity field we then include a deviation from a linear radial velocity component (Model 1) or a non-zero poloidal component according to Steffen, et al. (2009) (Model 2). We then make second-order corrections for the structure of the main shell to fit the outline of the HST images. After these, similar adjustments are made to the structure along the line of sight and in the velocity field in order to fit the small scale structure of the P-V diagrams. Adjustments are then made to small-scale features in the brightness distribution. This process is iterated until a satisfactory model is obtained.

A random distribution of particles is located either on the surface or in the volume of the mesh. They are the points where 3D space is sampled. The local average density of the particles is constant on the surface or in the volume distribution. The stochastic character of particle distribution introduces local noise in the brightness distribution which can be controlled by the number of particles used. With the appropriate particle density, this noise simulates the small-scale emissivity variations. Thus, the precise location of these variations is modeled only statistically. Some small-scale features have been included adding or deleting particles locally with a special tool for that purpose. The distribution of sampling points is the first spatial selection of the sub-structure that constitutes the whole object. A second selection is obtained via the assignment of a relative emissivity, which may be zero to exclude emission from certain mesh regions. The emissivity is given as a function of spatial coordinates. This function can either be analytic or set by an interactive graph. Similarly, the velocity vector is assigned as an analytical function of coordinates or an interactive graph in each coordinate (Steffen, et al. 2009).

The interactive graphical representation of these quantities allows for very detailed and complex structures to be reproduced. For instance, Figure (5) shows the relative emissivity distribution $e_{\theta}(\theta)$ on the equatorial ring structure as a function of azimuthal angle $\theta$ in spherical coordinates. The distribution in each coordinate is then multiplied to the distribution in other coordinates $e_r(r)$ and $e_{\phi}(\phi)$, i.e. $e(r,\phi,\theta) = e_0 \cdot e_{\theta}(\theta) \cdot e_r(r) \cdot e_{\phi}(\phi)$ (i.e. the distributions are assumed to be separable in their coordinates). $e_0$ is a total scaling factor.

After the observational parameters have been set (like orientation, spectrograph slit position and width, seeing and velocity resolution) the emissivity distribution is integrated along the line of sight and rendered as an image and PV diagram. In this model, we assume that the object is optically thin. Currently, the dynamic range of brightness modeling in {\em Shape} does not reach that of the features in NGC~7009, which in some cases are only clearly visible on a logarithmic scale. Since the precise relative brightness of large and small-scale features is not the main concern of this work, rather than the location and kinematics of structure in 3D space, this limitation is not important for our modeling. The modeling of relative brightness between features has been adjusted using only visual inspection. The images and PV diagrams have been displayed in a square root grayscale.

The comparison between model and observation is done by superimposing them with variable transparency directly in Shape. This allows a very accurate visual comparison of positions and outlines in the images and PV diagrams.

\begin{figure}[!t]\centering
  \vspace{0pt}
  \includegraphics[angle=-90,width=0.95\columnwidth]{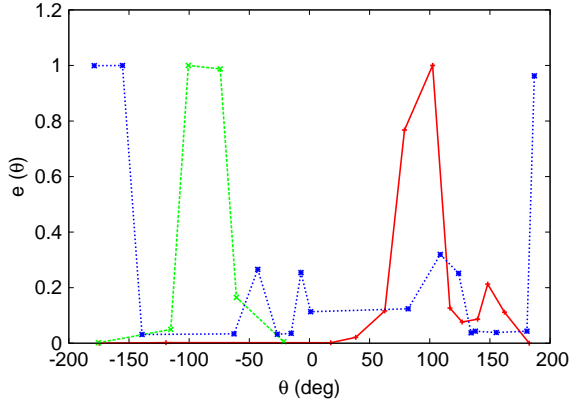}
  \caption{The relative emissivity distribution as a function of azimuthal angle $\theta$ around the
  axis of the object for the north-western (solid line) and south-eastern (dashed line) ring and the
  oblique equatorial ring (dotted line). Emissivities are normalized to their maximum value for each component.
   }
  \vspace{0pt}
  \end{figure}

\begin{figure}[!t]\centering
  \vspace{0pt}
  \includegraphics[angle=-0,width=0.95\columnwidth]{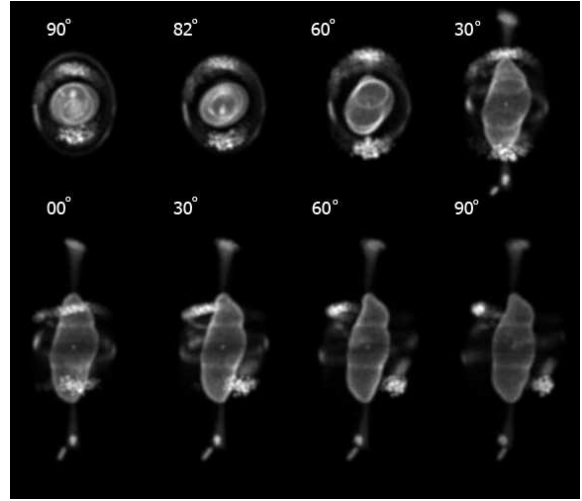}
  \caption{Images of the {\it Shape}-model as seen from selected viewpoints.
In the top row the observer's point of view is rotated around the
horizontal axis by the labeled angles.
In the bottom row the viewpoint is rotated around the vertical axis by
the specified angle. The spatial resolution in these images is 0.5 arcsec.
}
  \vspace{0pt}
  \end{figure}

For NGC~7009, all structures, except the main shell, have been modeled as volume distributions. The main shell is thought to be the thin, swept-up shell around the fast wind and is much thinner than its size. The outer shell is approximated as a filled triaxial ellipsoid that is limited on the inside by the main shell; this shell has been sampled between the outer ellipsoidal structure and the inner main shell. Other structures have been sampled throughout the volume that is set by their mesh. Note that this does not necessarily mean that emission comes from the full sampled region, because the relative emissivity distribution may limit the emission to only part of this volume.

\begin{figure}[!t]\centering
  \vspace{0pt}
  \includegraphics[angle=-90,width=0.95\columnwidth]{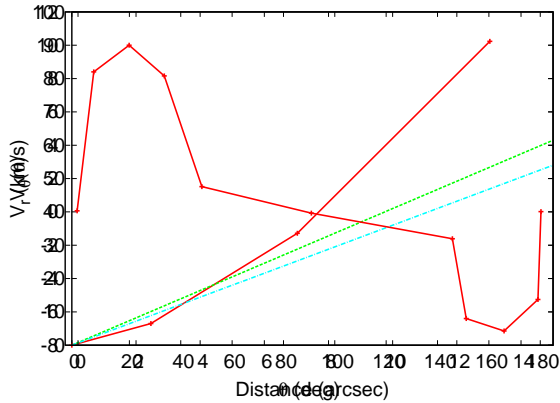}
  \caption{The radial velocity component is shown as a function of distance from
  the central star. The continuous line is for the main shell in Model 1.
  The dashed line is that of Model 2, which has a poloidal
  velocity component as show in Figure 7. The same radial velocity law (dashed line) has been adopted
  for the streams and ansae (4.1 km s$^{-1}$ arcsec$^{-1}$). The dot-dashed line is the velocity law adapted
  for the envelope, caps and equatorial rings (3.6 km s$^{-1}$ arcsec$^{-1}$).
   }
  \vspace{0pt}
  \end{figure}

\begin{figure}[!t]\centering
  \vspace{0pt}
  \includegraphics[angle=-90,width=0.95\columnwidth]{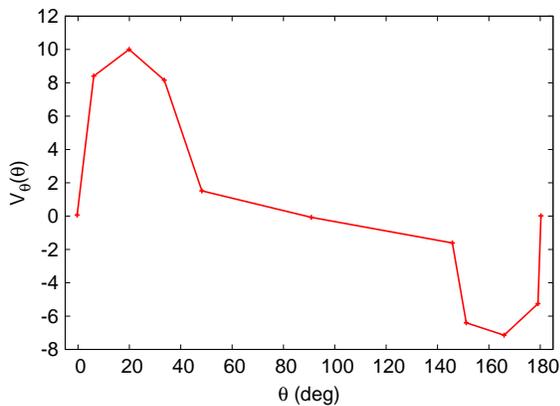}
  \caption{The poloidal velocity component of the main shell is shown as a function of latitude (with the equator at $90^o$. The eastern tip of the main shell is at $0^\circ$ and the western tip is at $180^\circ$.}
  \vspace{0pt}
  \end{figure}

\section{Results and discussion}

The full model structure of NGC~7009 is shown in Figure (6), where the viewpoint has been rotated around the horizontal axis (top row) and the vertical axis (bottom row). Model 1 and 2 have the same 3D structure, but different velocity fields, such that only one set of images has been shown. The position-velocity diagrams for both models can be compared to the observed data in Figures 3 and 4.

\subsection{Main shell}

We find that the data used in this work are consistent with a bent structure of the main shell, both in the plane of the sky and along the line of sight (Figure 6).

Model 1 has a purely radial velocity field with an outward increasing velocity gradient (Figure 7, solid line). Model 2 has a linear radial velocity component (Figure 7, dashed line) and a poloidal component as shown in Figure 8.

Since the purely radial velocity field is spherically symmetric, any differences between the western and eastern hemispheres in the PV diagrams have to be accounted for by structure along the line of sight. Although the increasing velocity gradient in Model 1 reduces the difference in the cross section of the main shell along the
line of sight compared to that in the plane of the sky (at high latitudes), the western tip is still much wider along the line of sight than in the plane of the sky. This is due to the fact that in the PV diagrams the western and the eastern tip have a similar thickness, but there is a considerable difference between the eastern and western tip in the image (while the radial velocity field is symmetric).

We have explored a model with the same radial velocity structure as Model 1, but with an unbent structure adjusting the cross section of the main shell along the line of sight. Excellent agreement with the PV diagram along the main axis can be achieved. However, this solution can not be brought into agreement with PV diagrams from intermediate position angles and has therefore been omitted. This implies that the bending is at an angle to the plane of the sky and the plane marked by
the main axis of the object and the line of sight.

The fact that X-ray emission has been observed that is constrained by the main shell (Guerrero, Gruendl \& Chu, 2002), makes it likely that it is still driven by the stellar wind or at least an over-pressured hot bubble. Following Steffen, et al. (2009), in Model 2 we have fitted a non-zero poloidal velocity field. Figure (8) shows the poloidal velocity as a function of latitude (with $90^o$ being the equator). In addition to the features in the PV diagrams, the main constraint for this velocity distribution was to have a main shell with a cross-section that was as circular as possible. As can be seen in Figure (6), top row, the cross section is non-circular. This is required by the structure of slits that are approximately perpendicular to the main axis of NGC~7009. The adopted relative emissivity distribution of the main shell is plotted in Figure (9).

In Figures (3) and (4) we compare Models 1 and 2 with the observed [NII] PV diagrams from STC04. Overall, the agreement with the data is very good. However, there are subtle deviations in Model 1 which cannot be eliminated unless traded for more serious deficiencies in other regions. In our opinion, these differences make Model 2 superior (which has a poloidal velocity component adding degrees of freedom) to Model 1 with a purely radial velocity field.  A comparison of the observations and model PV diagrams at mid to high latitudes (PA = 49$^o$ - 109$^o$) shows better agreement of Model 2 with the data, except for the lower (East-Southeast, E-SE) section at PA = 94$^o$. It has proved impossible to obtain a satisfactory agreement simultaneously between this E-SE region and the PV diagram along the main axis at PA = 79$^o$.

Regarding the accuracy of the velocity field, we estimate that, within our modeling framework and our basic assumptions about the cross section of the main shell, the graph for the upper limit of the radial velocity is accurate to about $10-15 \%$, whereas that of the poloidal velocity component is probably accurate to about $15-25 \%$. This has been estimated by varying the parameters manually until the model was clearly inconsistent with the observations. The main uncertainty is in the cross section of the object. For the combination of velocity structure and cross section there is no unique solution based on imaging and Doppler velocity measurements.

Considering the uncertainties, the differences in agreement with the observations are not sufficient
to definitely favor one model over the other. Considering the results from hydrodynamical simulations (Steffen et al., 2009), the likely solution is somewhere in between Models 1 and 2.
This is consistent with simulations that show a velocity field with, both, a radial component with an increasing gradient and a poloidal velocity component. Further restriction will have to wait for detailed internal proper motion measurements of individual features in the main shell. In Figure 10 we therefore make a prediction of the proper motion pattern expected for Model 2. A statistical analysis of such a proper motion pattern in an observed object will provide the observed equivalent of the graph in Figure 8. We expect all measured poloidal velocity values to fall below those of Model 2, with significant values beginning at a latitude of about 40$^o$. The outline of the highest values in the observed graph will directly provide the corresponding model component. This will then allow an observational derivation of the radial component as a function of distance from the central star (Figure 7).

Although we have not made a very detailed analysis of relative brightness of features in NGC~7009, we still find some interesting results for the main shell which are related to its structure. The relative model emissivity distributions of the main shell as a function of distance along the symmetry axis and as a function of the angle around the axis are shown in Figure (9) (note that the ratios between peaks is suppressed in our model, due to the reduced dynamical range as compared to NGC~7009). As a function of distance along the axis the two halves are very similar. But we find that the western and eastern half are brightest at approximately opposite sides almost along the line of sight (Figure 9, bottom, $\theta = 0$ is in the plane of the sky). This brightness difference might well be related to the bent structure of the main shell and the brightness of the caps at similar angles around the axis (see Figure 6).

Considering these results, the bent and point-symmetric structure together with the increased velocity deviating from homologous expansion (either radial or poloidal) at high latitudes suggest the existence of two separate mass-loss events that were at least partially collimated, but in different directions.

Because of the presence of a hot bubble, we have investigated the physically possible case of a velocity that is locally perpendicular to the main shell with increasing magnitude with distance from the star. We found no plausible structure that would fit the observations at high latitudes. Such a model could only be fitted to the data at latitudes of less than approximately $60^{\circ}$ from the equator.

\begin{figure}[!t]\centering
  \vspace{0pt}
  \includegraphics[angle=-90,width=0.95\columnwidth]{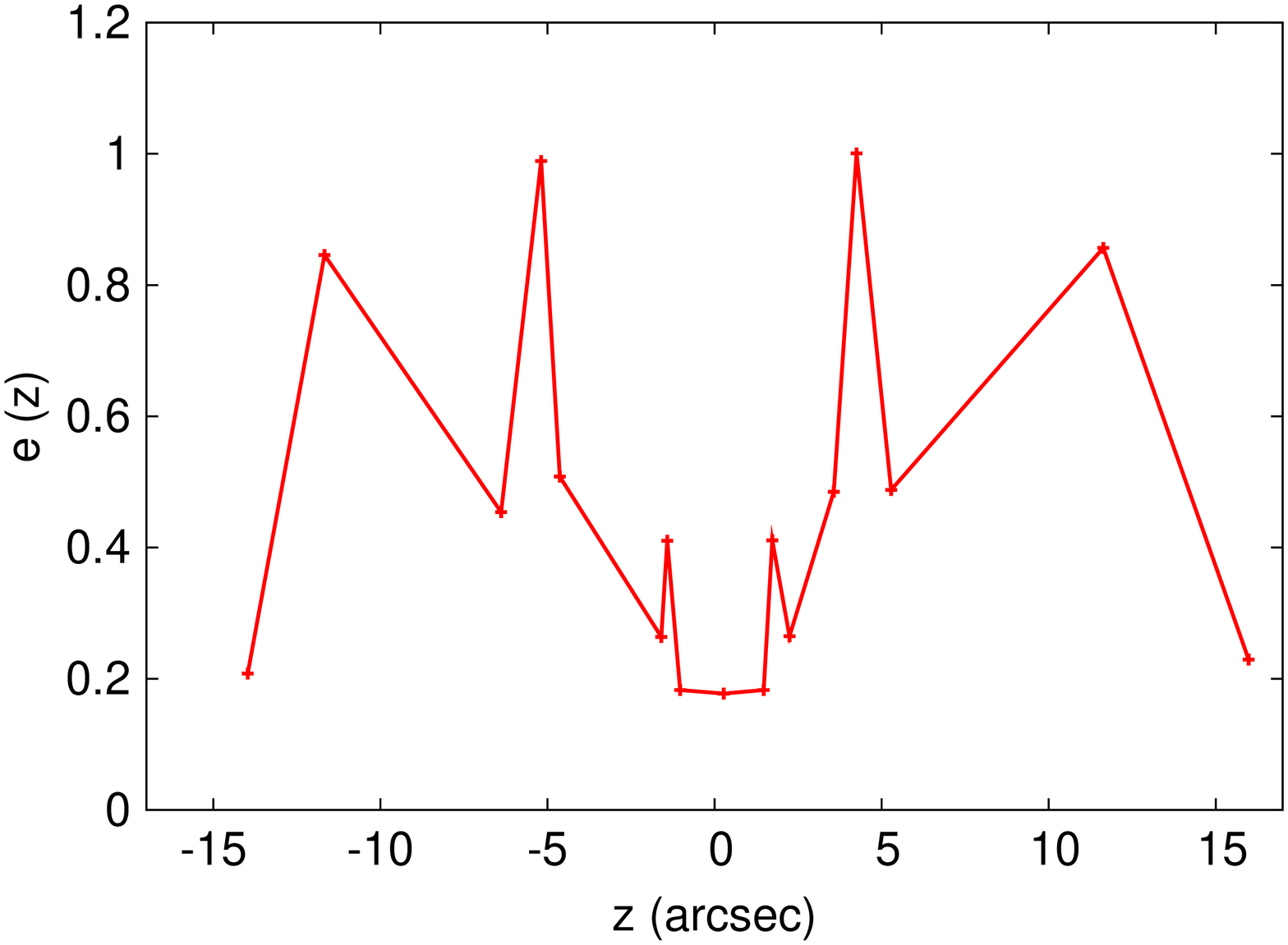}
  \includegraphics[angle=-90,width=0.95\columnwidth]{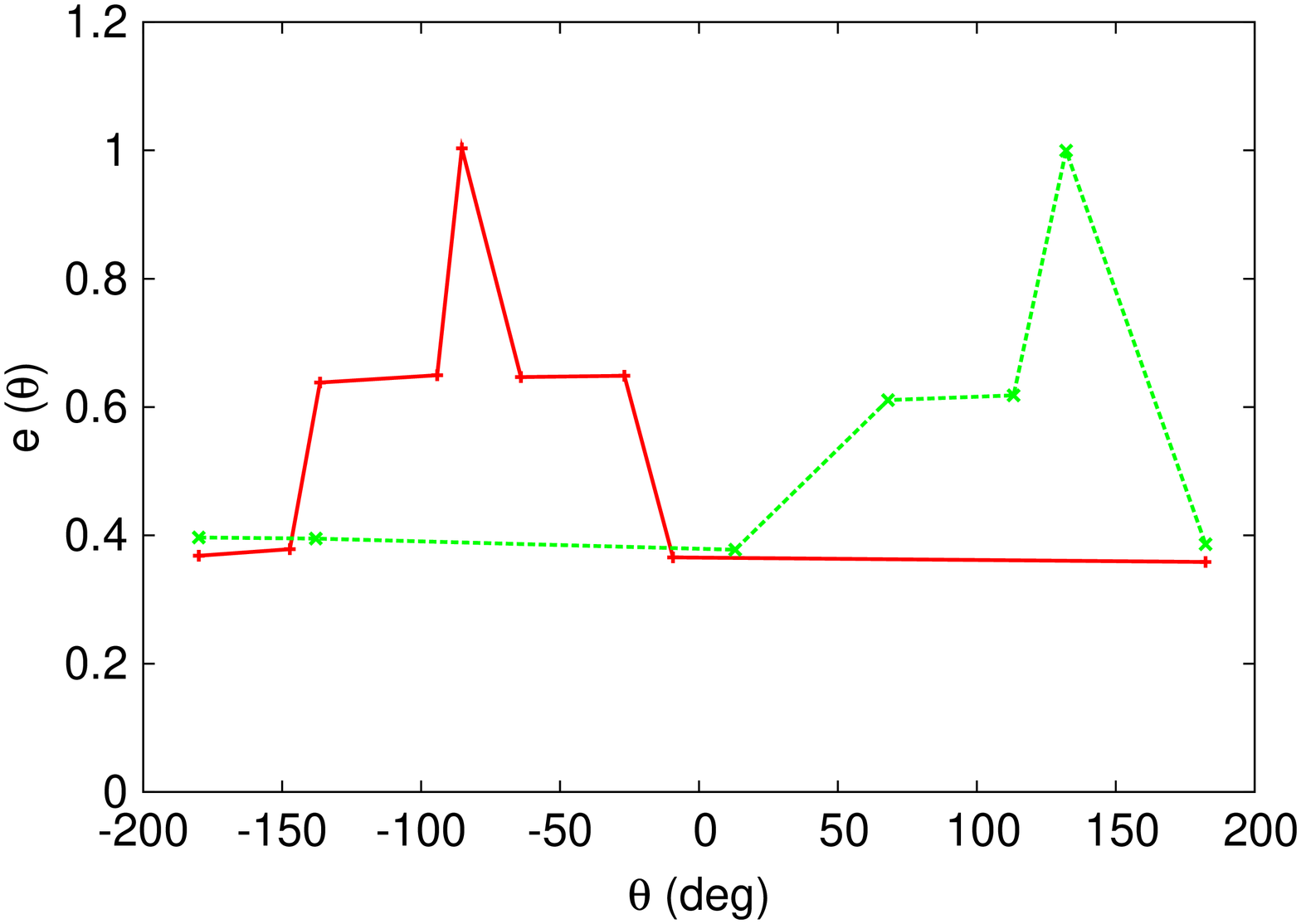}
  \caption{The model's relative emissivity of the main shell is plotted as a function of
  the distance along the axis $z$ (top) and as a function of azimuthal angle $\theta$ (bottom).
  The distribution in angle is different for the northern (continuous line) and the southern
  half (dashed line) with peaks at nearly opposite angles. Negative z-values in the top graph
  are toward the East.}
  \label{fig9}
  \vspace{0pt}
  \end{figure}

\subsection{Ansae}

The main question about the ansae is their formation process. Did they originate near the
central star, or are they the result of a global collimation process similar to a hydrodynamic Cant\'o flow (Cant\'o, J., Tenorio-Tagle, G., R\'o\.zyczka, M., 1988) or magnetic focusing (\ggs \& L\'opez, 2000).
Velocity vectors that point directly to the center would speak in favor of an origin near the central star, whereas vectors that converge near the exit point from the main shell might be regarded as evidence for global focussing.

Rodr\'{\i}guez \& G\'omez (2007) use Very Large Array radio observations to determine the proper motion of the ansae. They find $23\pm6$ and $34\pm10$ mas yr$^{-1}$ for the eastern and the western ansae, respectively, i.e. the western ansa is almost 1.5 times faster than the eastern one. However, within the uncertainty, they could
also be the same. We have therefore tested both options. The results shown in all figures refer to the case of equal velocities using the 4.1 km s$^{-1}$ arcsec$^{-1}$ velocity law shown in Figure (7).

Regarding the origin of the ansae, the measured difference in velocity might be quite important. The width of the ansae in Doppler velocity (see Figure 3, PV diagram at PA = 79$^o$) as compared to the width of the main shell is larger than the ratio of widths in the images, especially for the eastern ansa. This may be, of course, simply because the ansae are actually more extended along the line of sight (see the reconstruction in STC04). Let us assume that the size of the ansae is approximately the same along the line of sight as in the plane of the sky, then the result is quite different. If, in addition, the velocities of both ansae are at the lower end of the measured values, then their Doppler velocity width requires that the velocity vectors focus near the tips of the main shell (see Figure 10). If the western ansa actually has a velocity that is 1.5 times higher than the eastern counterpart, then
the spread in the PV diagram is consistent with velocity vectors emerging from the central star. However, this is true only for the western ansa, whereas the eastern ansa definitely requires velocity vectors that expand from a position near the tip of the main shell. In order to further test the origin of the ansae, high-resolution observations of the proper motion of individual knots in the ansae will be needed. In the [NII] the knots are so well defined, that such a study might be possible using the HST within a few years.

\subsection{Caps and rings}

For completeness we have included caps and rings in our model, following the same homologous expansion law for all features (3.6 km s$^{-1}$ arcsec$^{-1}$, see Figure 7).
We confirm the existence of at least one, somewhat oblique, equatorial ring. Additional ring segments at different inclination angles to the equator have been included to model some of the small-scale features in the PV diagrams and the emission. Their relative emissivity of the main equatorial ring as a function of angle around the main axis has been plotted in Figure (5). The same figure also shows the relative brightness distribution used to model the caps as regions in the rings at some distance from the equator (solid and dashed line; see also the mesh in Figure 2). Note that the angular position of the bright caps is similar to that of the bright regions in the main shell (see Figure 9). The caps have been modeled with few particles in order to reproduce their small scale structure. However, since the particles have been distributed randomly, knots in the observations have no particular corresponding counterpart in the model.

\subsection{Proper motion pattern}

Figure (10) shows the proper motion pattern predicted from Model 2. The projected velocity vectors have been drawn for a small random subset of particles in the model. Since there are several velocity subsystems, observation of the proper motion pattern requires local methods, rather than global expansion measurements that might be described by a single expansion factor (Fern\'andez, Monteiro \& Schwarz, 2004). It is preferable to apply methods based on the motion of individual features like those used by Li, Harrington \& Borkowski (2002) for the young planetary nebula BD+30$^\circ$3639. Similar methods have been employed on NGC~6302 by Meaburn et al. (2005) and by O'Dell et al. (2009) on NGC~6720, the Ring Nebula.
For clarity, we have included a few guiding lines that emphasize the non-radial directions of proper motion in the main shell and the ansae. In an observational test, the poloidal velocity component could be derived from the proper motion vectors at the edge of the nebula. Plotted as a function of latitude, such results could be directly compared with the upper limits for the poloidal velocity components predicted in Figure (8).

\begin{figure*}[!t]\centering
  \vspace{0pt}
 \includegraphics[angle=00,width=1.95\columnwidth]{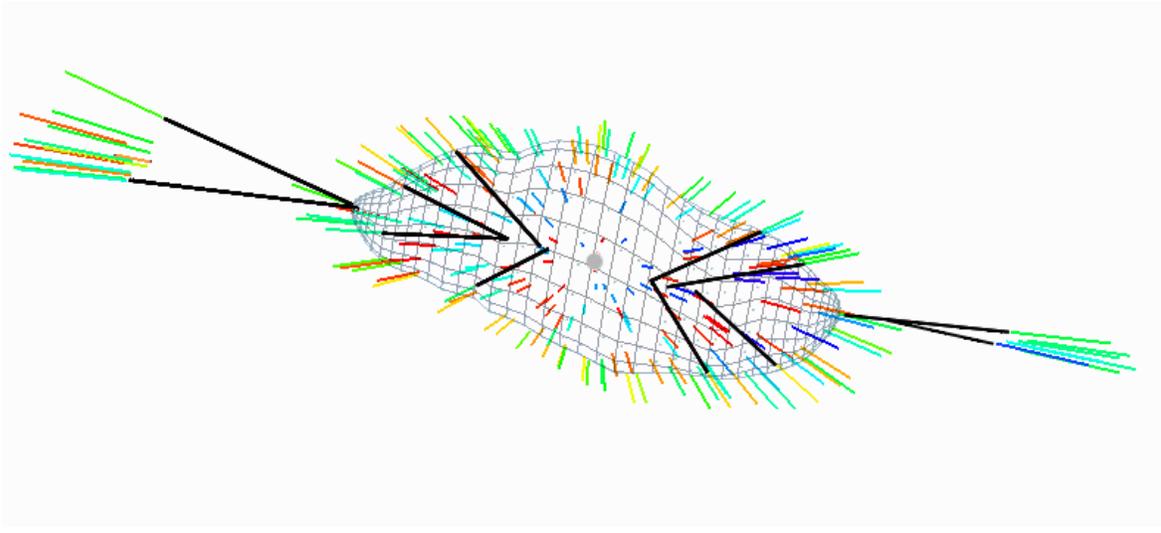}
  \caption{The projected velocity vectors for a randomly selected subset of
particles is shown. They represent
the proper motion pattern predicted with the {\it Shape}-model. The
color coding is according to the
Doppler-shift of the particle. In order to show the off-center convergence of the
proper motion vectors at mid-latitudes and the ansae,
extended guiding lines have been drawn which extend some of the
proper motion vectors backwards.
To assist in locating the positions of the vectors in the image of
NGC~7009, the model mesh of the main
shell has been included. The position of the cental star has been marked with a filled circle.}
  \vspace{0pt}
  \end{figure*}

In order to measure the expansion of NGC~7009, Fern\'andez, Monteiro \& Schwarz (2004) have shown difference images based on [NII] HST images from 1996 and 2001. Unfortunately, one of the HST images did not include the western ansa. They also had some problems appropriately registering the two images, such that there might be some systematic errors in their results. It is therefore of interest to predict future results of such measurements based on our 3D model.
Fern\'andez, Monteiro \& Schwarz (2004) have not analyzed the expansion of the main shell, arguing that it might be a moving ionization front, which does not provide the velocity of the bulk motion of the gas. However, Sabbadin et al. (2004) find that the nebula is fully ionized, possibly except in the ansae and the caps. Therefore, whatever causes the observed expansion, a study of this expansion might bring new insights. We therefore predict the expected expansion pattern for Model 2, based on the assumption that the expansion is due to bulk motion.

{\it Shape} has the option to use the model velocity field to advance the position of the individual particles in time, assuming a velocity that does not change with time. With this feature we produced a difference image for various time-spans.
Figure (11) shows the expected difference between [NII] images taken 50 years apart assuming a distance of 1.6 kpc (Sabbadin et al., 2004). Note that the distance values given for this object vary approximately between  0.8 and 1.6 kpc. At half the distance the same expansion would be obtained in 25 years.
Smaller time-scales provide similar, but noisier results. Since the spatial resolution of the HST images is better than that of our model by about a factor of three, a similar difference image with HST resolution could be obtained with a baseline of 15-20 years. A new observational analysis of this type is of interest, since there is a systematic difference between the results from our model and those of Fern\'andez, Monteiro \& Schwarz (2004). In their images, the eastern tip of the main shell clearly shows stronger expansion north of the symmetry axis. In our model, there is no such systematic difference between North and South.

\begin{figure}[!t]\centering
  \vspace{0pt}
  \includegraphics[angle=00,width=0.95\columnwidth]{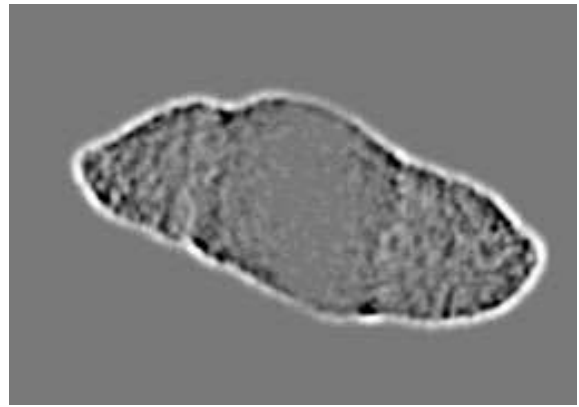}
  \caption{ The difference image of the model for [NII] emission for time 1996 and 50 years later is shown. Zero emission was normalized to 128 on a grey-scale of 255 shades. The dark regions show the residuals of the older image, whereas the bright residuals are due to the future image.
}
  \vspace{0pt}
  \end{figure}

\section{Conclusions}

Based on detailed position-velocity diagrams from Sabbadin et al. (2004) and Hubble Space Telescope imagery, we have constructed a high-resolution 3D morpho-kinematical model for the planetary nebula NGC~7009. We largely confirm the results by Sabbadin et al. (2004), but obtain differences for the main shell due to the differences in the assumptions on the velocity fields. We have estimated the upper limits for deviations from a homologous expansion. We find that the data are consistent with a radial velocity field with a stronger gradient at high latitudes (Model 1) as compared to the region near the equator. In a second model we have assumed a linearly increasing radial velocity component with an added poloidal component of order 10~km/s at latitudes of about $70^\circ$. Based on hydrodynamical simulations of wind-driven PNs by Steffen et al. (2009), we expect these velocity fields to represent limiting cases and that the real nebula has a velocity field having a combination of increasing gradient in the radial velocity and a poloidal component, but each with smaller values. We exclude the possibility that the velocity field of the main shell is locally perpendicular to its surface at least at high latitudes. We also find that the expansion of the ansae is non-radial if the central star is used for reference. Their velocity field is focused near the apparent exit points from the main shell. This points towards an origin as a Cant\'o flow, large scale magnetic focusing or direct interaction with the ambient medium. High-resolution proper motion measurements of individual features in the main shell and the ansae could yield a test of the non-radial expansion patterns by now or within a few years. We predict the proper motion pattern for the model with a non-zero poloidal velocity component.

\section{Acknowledgements}
We acknowledge support from CONACYT grant 49447 and UNAM DGAPA-PAPIIT  IN108506-2. M.E. and S.M. acknowledge support by the {\it Instituto de Astronom\'{\i}a, Universidad Nacional Aut\'onoma de M\'exico}. N.K. acknowledges support by the Natural Sciences and Engineering Research Council of Canada (NSERC) and the Killam Trusts. We thank Franco Sabbadin and collaborators for kindly granting permission to adapt their images and PV diagrams of NGC~7009 for this work.

\end{document}